\documentstyle[11pt,emulateapj,apjfonts,epsf]{article}

\submitted{Accepted for publication in The Astronomical Journal}

\righthead{METCALFE}
\lefthead{GENETIC-ALGORITHM-BASED OPTIMIZATION APPLIED TO BH~CAS}

\begin{document}

\title{Genetic-Algorithm-based Light Curve Optimization\\
       Applied to Observations of the W~UM\lowercase{a} star BH~Cas}

\author{Travis S. Metcalfe\altaffilmark{1}}

\affil{Department of Astronomy, University of Texas-Austin}
\altaffiltext{1}{formerly at the Department of Astronomy, University of Arizona}
\authoraddr{Mail Code C1400, Austin, TX 78712; travis@astro.as.utexas.edu}

\begin{abstract}

I have developed a procedure utilizing a Genetic-Algorithm-based
optimization scheme to fit the observed light curves of an eclipsing
binary star with a model produced by the Wilson-Devinney code. The
principal advantages of this approach are the global search capability and
the objectivity of the final result. Although this method can be more
efficient than some other comparably global search techniques, the
computational requirements of the code are still considerable. I have
applied this fitting procedure to my observations of the W~UMa type
eclipsing binary BH~Cassiopeiae. An analysis of V--band CCD data obtained
in 1994/95 from Steward Observatory and U-- and B--band photoelectric data
obtained in 1996 from McDonald Observatory provided three complete light
curves to constrain the fit. In addition, radial velocity curves obtained
in 1997 from McDonald Observatory provided a direct measurement of the
system mass ratio to restrict the search. The results of the GA-based fit
are in excellent agreement with the final orbital solution obtained with
the standard differential corrections procedure in the Wilson-Devinney
code.

\end{abstract}


\section{Introduction}

The problem of extracting useful information from a set of observational
data often reduces to finding the set of parameters for some theoretical
model which results in the closest match to the observations. If the
constitutive physics of the model are both accurate and complete, then the
values of the parameters for the `best-fit' model can yield important
insights into the nature of the object under investigation.

When searching for the `best-fit' set of parameters, the most fundamental
consideration is: where to begin? Models of all but the simplest physical
systems are typically non-linear, so finding the least-squares fit to the
data requires an initial guess for each parameter. Generally, some
iterative procedure is used to improve upon this first guess in order to
find the model with the absolute minimum residuals in the
multi-dimensional parameter space.

There are at least two potential problems with this standard approach to
model fitting. The initial set of parameters is typically determined by
drawing upon the past experience of the person who is fitting the model.
This {\it subjective} method is particularly disturbing when combined with
a {\it local} approach to iterative improvement.  Many optimization
schemes, such as differential corrections (Proctor \& Linnell \markcite{a}
1972) or the simplex method (Kallrath \& Linnell\markcite{b} 1987), yield
final results which depend to some extent on the initial guesses. The
consequences of this sort of behavior are not serious if the parameter
space is well behaved---that is, if it contains a single, well defined
minimum. If the parameter space contains many local minima, then it can be
more difficult for the traditional approach to find the global minimum.

\section{Genetic Algorithms}

An optimization scheme based on a Genetic Algorithm (GA) offers an
alternative to more traditional approaches. Restrictions on the range of
the parameter space are imposed only by observations and by the physics of
the model. Although the parameter space so defined is often quite large,
the GA provides a relatively efficient means of sampling globally while
searching for the model which results in the absolute minimum variance
when compared to the observational data. While it is difficult for GAs to
find precise values for the set of `best-fit' parameters, they are well
suited to search for the {\it region} of parameter space that contains the
global minimum. In this sense, the GA is an objective means of obtaining a
good first guess for a more traditional method which can narrow in on the
precise values and uncertainties of the `best-fit'.

The underlying ideas for Genetic Algorithms were inspired by Darwin's
\markcite{01}(1859) notion of biological evolution through natural
selection. A comprehensive description of how to incorporate these ideas
in a computational setting was written by Goldberg\markcite{02} (1989). In
the first chapter of his book, Goldberg describes the implementation of a
simple GA---involving several steps which are analogous to the process of
biological evolution.

The first step is to fill the parameter space uniformly with trial
parameter-sets which consist of randomly chosen values for each parameter.
The theoretical model is evaluated for each trial parameter-set, and the
result is compared to the observational data and assigned a {\it fitness}
which is inversely proportional to the root-mean-square residuals. The
fitness of each trial parameter-set is mapped into a survival probability
by normalizing to the highest fitness. A new generation of trial
parameter-sets is then obtained by selecting from this population at
random, weighted by the survival probabilities.

Before any manipulation of the new generation of trial parameter-sets is
possible, their characteristics must be encoded in some manner. The most
straightforward way of encoding the parameter-sets is to convert the
numerical values of the parameters into a long string of numbers. This
string is analogous to a chromosome, and each number represents a gene.
For example, a two parameter trial with numerical values $x_1=1.234$ and
$y_1=5.678$ would be encoded into a single string of numbers `12345678'.

The next step is to pair up the encoded parameter-sets and modify them in
order to explore new regions of parameter space. Without this step, the
final solution could ultimately be no better than the single best trial
contained in the initial population. The two basic operations are {\it
crossover} which emulates reproduction, and {\it mutation}.

Suppose that the encoded trial parameter-set above is paired up with
another trial having $x_2=2.468$ and $y_2=3.579$ which encodes to the
string `24683579'. The single-point crossover procedure chooses a random
position between two numbers along the string, and swaps the two strings
from that position to the end. So if the third position is chosen, the
strings become
$$
\begin{array}{c}
123|45678 \rightarrow 123|83579 \\
246|83579 \rightarrow 246|45678
\end{array}
$$
Although there is a high probability of crossover, this operation is not
applied to all of the pairs. This helps keep favorable characteristics
from being eliminated or corrupted too hastily. To this same end, the rate
of mutation is assigned a relatively low probability. This operation
allows for the spontaneous transformation of any particular position on
the string into a new randomly chosen value. So if the mutation operation
were applied to the sixth position of the second trial, the result might
be
$$
24645|6|78 \rightarrow 24645|0|78
$$

After these operations have been applied, the strings are decoded back
into sets of numerical values for the parameters. In this example, the new
first string `12383579' becomes $x_1=1.238$ and $y_1=3.579$ and the new
second string `24645078' becomes $x_2=2.464$ and $y_2=5.078$. Obviously,
the new set of trial parameter-sets is related to the original set in a
very non-linear way. This new generation replaces the old one, and the
process begins again: the model is evaluated for each trial, fitnesses are
assigned, and a new generation is constructed from the old and modified by
the crossover and mutation operations. Eventually, after a modest number
of generations, some region of parameter space remains populated with
trial parameter-sets, while other regions are essentially empty.  The
robustness of the solution can be established by running the GA several
times with different random number sequences.

Genetic Algorithms have been used a great deal for optimization problems
in other fields, but until recently they have not attracted much attention
in astronomy. The application of GAs to problems of astronomical interest
was promoted by Charbonneau\markcite{03} (1995), who demonstrated the
technique by fitting the rotation curves of galaxies, a multiply periodic
signal, and a magneto-hydrodynamic wind model. Many other applications of
GAs to astronomical problems have appeared in the recent literature.
Hakala \markcite{04}(1995) optimized the accretion stream map of an
eclipsing polar. Lang\markcite{05} (1995) developed an optimum set of
image selection criteria for detecting high-energy gamma rays.  Kennelly
{\it et al.}\markcite{06} (1996) used radial velocity observations to
identify the oscillation modes of a $\delta$ Scuti star. Lazio
\markcite{07}(1997) searched pulsar timing signals for the signatures of
planetary companions.  Charbonneau {\it et al.}\markcite{08} (1998)
performed a helioseismic inversion to constrain solar core rotation. Most
recently, Wahde \markcite{09}(1998)  used a GA to determine the orbital
parameters of interacting galaxies. The applicability of GAs to such a
wide range of astronomical problems is a testament to their versatility.

\section{Observations}

I have applied a GA-based optimization scheme to my observations of the
W~UMa type eclipsing binary star BH~Cassiopeiae. Due to an unfortunate
historical accident, this relatively bright object ($m_V=12.6$) was
neglected observationally for more than half a century prior to this
investigation.

\subsection{Background}

The discovery observations of BH~Cas were made by S.~Beljawsky from the
Sime{\"{\i}}s Observatory between June and September of 1928, and were
later reported in the paper ``34 New Variable Stars (Fourth Series)''
(Beljawsky\markcite{10} 1931). The eleventh variable in his list was in
the position of the star which later came to be known as BH~Cas. Beljawsky
gave it the temporary designation `353.1931 Cass' at the time of
discovery. No maximum or minimum magnitudes were listed.

Follow-up observations of BH~Cas were made by B.~Kukarkin with the 180-mm
Zeiss-Heyde refractor from the Moscow Observatory and/or the 180-mm
Grubb-Mart refractor from the Tashkent Observatory. Kukarkin began
gathering observations in April 1936, and continued through the summer of
1937. The results of his observations were reported in the paper
``Provisional Results of the Investigation of 80 Variable Stars in Fields
6, 13, 15 and 62.'' published in the journal {\it Ver\"anderliche Sterne
Forschungs-und Informationsbulletin} (Kukarkin\markcite{11} 1938). Based
on his 62 visual observations and an examination of the plate collection
at the Moscow Observatory, Kukarkin concluded that BH~Cas was possibly
W~UMa type with a period near 0.5 days and an amplitude of 0.4 magnitudes.

The most recent edition of the {\it General Catalogue of Variable Stars}
(Kukarkin {\it et al.}\markcite{12} 1969) contains a footnote on the entry
for BH~Cas which refers to a 1943 paper by P.~Ahnert and C.~Hoffmeister.
After searching for BH~Cas on photographic plates, and visually with a
350-mm reflector, they concluded that ``no star in the surroundings of the
indicated place shows light variation...'' (Ahnert \& Hoffmeister
\markcite{13}1943). After the appearance of this article, no observations
of BH~Cas appeared in the literature for more than 50 years.  Considering
the strength of their conclusion, the absence of additional efforts to
observe this object is not surprising.

On the night of 19 February 1994, I obtained a series of CCD images of the
region centered on BH~Cas over a period of $\sim$1.3 hours using the
Spacewatch CCD at the Steward Observatory 0.91-m telescope on Kitt Peak.
For each image, I measured the flux from BH~Cas and from the comparison
and check stars, GSC~01629 and GSC~01134 respectively.  Plots of the
relative flux over time revealed an increase in the brightness of BH~Cas
relative to the comparison star while the latter remained constant
relative to the check star (Metcalfe\markcite{14} 1994).

\subsection{Photometry}

On twelve nights between September 1994 and October 1995 I used the
`2kBig' CCD at the Steward Observatory 1.5-m telescope to obtain images of
BH~Cas and the surrounding area for photometric study. These observations
allowed me to reconstruct a complete light curve in the V--band from a
total of 432 data points, as well as partial curves in the R-- and
I--bands.

For each night of observations, I reconstructed a flat-field image from a
median combination of many data frames with non-overlapping star images. I
used the IRAF {\it ccdproc} package to clean and calibrate each image, and
the {\it phot} package to extract aperture photometry for BH~Cas, the
comparison star GSC~00784, and the anonymous check star (see Figure 1).

For ten nights in December 1996 I used the `P3Mudgee' 3-channel 
photoelectric photometer (Kleinman, Nather \& Phillips
\markcite{15}1996) ~on ~the ~2.1-meter ~telescope ~at ~McDon-


\epsfxsize 3.5in
\epsffile{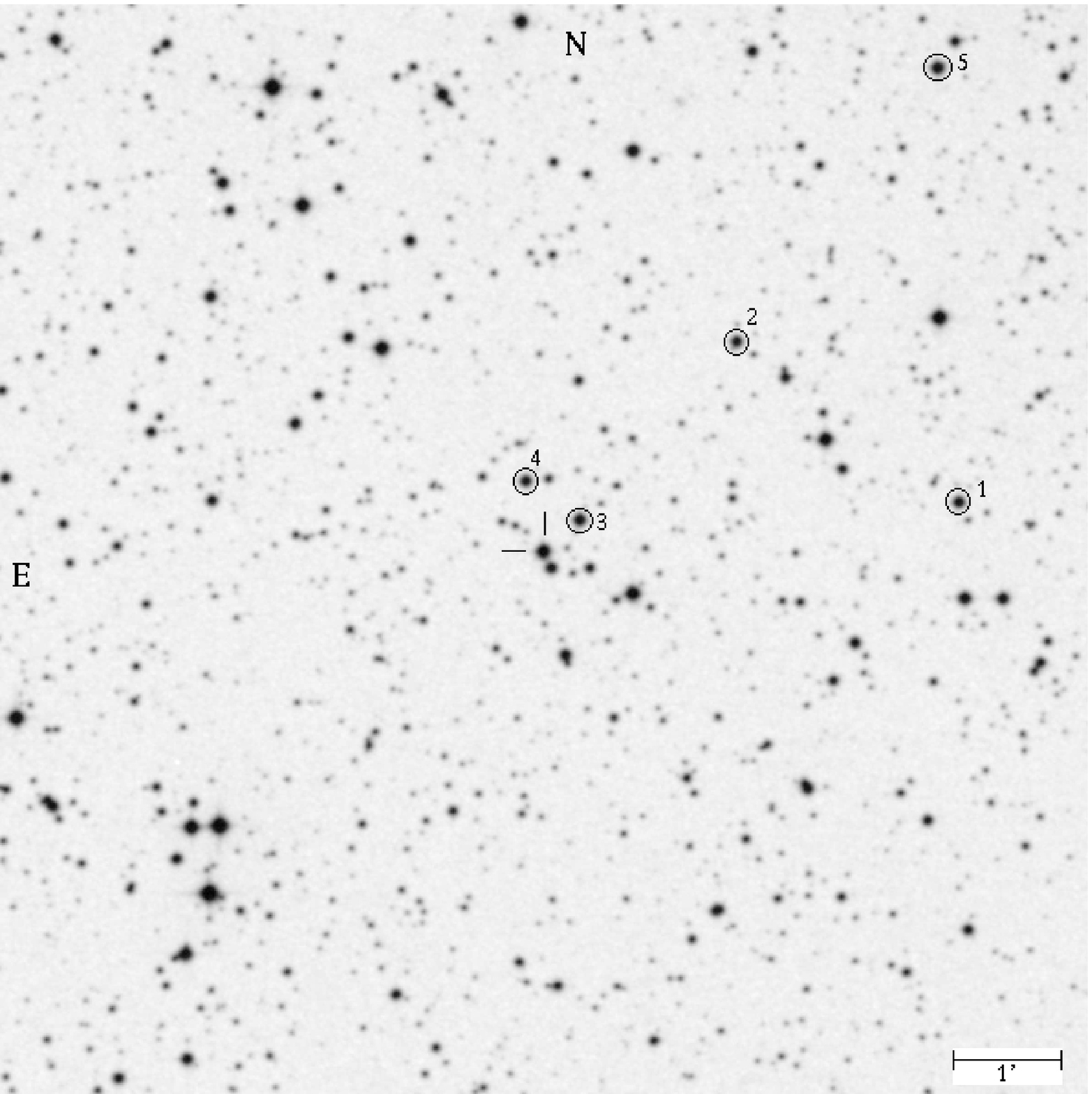}
\vskip 10pt
\noindent{\sc Fig.~1.---} A $10\arcmin\times10\arcmin$ image from the
Digitized Sky Survey centered on BH~Cassiopeiae. Comparison and check
stars that were used in this study are circled: (1)~GSC~01629,
(2)~GSC~01134, (3)~GSC~00784, (4) anonymous check star for CCD data, 
and (5)~GSC~00594.\newline


\noindent ald Observatory to obtain multi-color observations of BH~Cas. 
These observations yielded complete U-- and B--band light curves with 1041 
and 1107 data points respectively. 


\begin{table*}
\tablenum{1}\label{tab1}
\centerline{TABLE 1}
\centerline{\sc Characteristics of Photometry Data Sets}
\begin{center}
\vskip -5pt
\begin{tabular}{lccccc}\hline\hline
Telescope           & Julian Day & Airmass     & Filter& $k'$ 	& $k''$	\\
\hline
Steward 1.5-m  & 2449767 & 1.54--2.61	& V	& 0.133	& \nodata\\
Steward 1.5-m  & 2449970 & 1.12--1.26	& V     & 0.079	& \nodata\\
Steward 1.5-m  & 2449971 & 1.12--1.21	& V     & 0.550	& \nodata\\
Steward 1.5-m  & 2449977 & 1.12--1.43	& V     & 0.362	& \nodata\\
Steward 1.5-m  & 2449978 & 1.12--1.40	& V     & 0.323	& \nodata\\
Steward 1.5-m  & 2449998 & 1.12--1.19	& V     & 0.147	& \nodata\\
McDonald 2.1-m & 2450430 & 1.14--2.81	& B	& 0.253	& 0.05\tablenotemark{a}\\
McDonald 2.1-m & 2450431 & 1.14--1.81	& B	& 0.230	& 0.05\tablenotemark{a}\\
McDonald 2.1-m & 2450430 & 1.14--2.81	& U	& 1.254	& 0.72\tablenotemark{a}\\
McDonald 2.1-m & 2450431 & 1.14--1.81	& U	& 0.775	& 0.33\tablenotemark{a}\\
\hline
\end{tabular}
\vskip 5pt
\hskip -3.1in {$^a$ Adopted value}
\end{center}
\end{table*}


I followed the standard reduction procedure (Clemens\markcite{16} 1993),
but some additional corrections were necessary.  Intermittent problems
with the filter wheel on all but the final night of observations required
that I make small ($\sim$3\%) normalization corrections derived from
nightly phototube cross calibrations. The considerable color difference
between BH~Cas and the comparison star GSC~00594 ($\Delta B-V\sim0.37$)
yielded significant second-order extinction for observations obtained at
high airmass. I adopted reasonable values for the second order extinction
coefficients based on a study by Kim \& Park\markcite{17} (1993). In this
paper, the authors demonstrated that the values of the extinction
coefficients can change drastically from one night to the next, and that
the `second-order' coefficient could be anywhere from negligible to
dominant.  Based on their Table 2, I determined that the median value of
the second-order coefficient in the B--band was 0.05. Adopting this value,
the resulting corrections applied to the B--band light curve of BH~Cas
typically amounted to a few hundredths of a magnitude. Although the paper
did not determine values of the second-order coefficient in the U--band,
it demonstrated the large range of acceptable values relative to the
first-order coefficient. This allowed me to adopt reasonable U--band
coefficients based on the assumption that the corrected maxima should be
approximately equal. The application of these corrections makes it
impossible to determine reliable estimates of the O'Connell effect
(Davidge \& Milone\markcite{18} 1984) from these data. Table \ref{tab1}
lists the airmass range and extinction coefficients for all of the data
sets used in this analysis.

\subsection{Spectroscopy}

On seven nights in September/October 1997 I obtained a time series of
spectra of BH~Cas using the Sandiford cassegrain echelle spectrograph
(McCarthy {\it et al.}\markcite{19} 1993) on the 2.1-m telescope at
McDonald Observatory. I adjusted the cross disperser and grating rotation
to provide wavelength coverage from 5430 to 6670 \AA. The velocity
resolution of this setup was $\sim$2 km s$^{-1}$.

I followed the standard IRAF reduction procedure for echelle spectra
(Churchill\markcite{20} 1995). Although most spectral orders contained at
least one weak metal line, there were only two strong features: the Na~D
lines and H$\alpha$. The Na~D lines were contaminated by narrow
atmospheric and interstellar features, so the only line which offered the
possibility of measuring reliable radial velocities was H$\alpha$.

The velocities were derived from cross correlations of the observed
spectra with a spectrum of the radial velocity standard star 31~Aql
obtained on each night ($v_r = 100.5$ km~s$^{-1}$; Astronomical
Almanac\markcite{AA} 1997). I determined the systematic stability of the
instrument and reduction procedure by cross correlating the spectra of
31~Aql obtained on different nights with each other. Systematic velocity
shifts between the data sets ranged from 2--10~km~s$^{-1}$ with
uncertainties in the range 1--3~km~s$^{-1}$. I corrected each of the
derived velocities for these small nightly offsets.

\section{Model Fitting}

\subsection{Wilson-Devinney Code} 

Before about 1970, the standard approach to modeling close binary stars
was that of Russell \& Merill\markcite{21} (1952). This {\it geometrical}
model revolved a system of two similar ellipsoids to produce a light
curve. While it was admittedly only a useful approximation of a true
binary system, this model could be treated analytically, which was
appropriate for the tools available at the time.

After a detailed treatment of close binary stars by Kopal\markcite{22}
(1959) which described the Roche equipotential surfaces, several authors
attempted to calculate light curves by revolving this {\it physical}
model. Progress was limited by the computational facilities that were
available at the time. Early attempts by Lucy\markcite{23} (1968) and Hill
\& Hutchings\markcite{24} (1970) represented substantial improvements over
the Russell model, but they were still incomplete.  Wilson \& Devinney
\markcite{25}(1971) introduced the model which has served as the
foundation for many improvements over the past few decades (Wilson
\markcite{26}1994).  At present, the majority of published results for
close binary stars are obtained using the Wilson-Devinney code
(Milone\markcite{27} 1993).

There are two modes of operation in the Wilson-Devinney (\hbox{W-D}) code for
overcontact binaries like W~UMa systems. One of these modes uses a single
continuous gravity darkening law for the entire common envelope to fix the
temperature of the secondary star. The other mode keeps the secondary
temperature a free parameter, allowing the model to be in physical contact
without being in thermal contact. Both overcontact modes require that the
two stars have identical surface potentials, gravity darkening exponents,
bolometric albedos, and limb darkening coefficients. Since the primary and
secondary eclipses in the observed light curves of BH~Cas have
significantly different depths, I chose to use the overcontact mode with
an adjustable secondary temperature. I used a blackbody radiation law and
simple reflection.

Some parameters of the model may be fixed initially from theory.  I
interpolated the values for the monochromatic limb darkening coefficients
from the tables in Al~Naimiy\markcite{28} (1978) to the proper temperature
and wavelength. I used the effective wavelengths of the UBV bandpasses and
the temperature derived from the colors of BH~Cas. The bolometric albedo
(also known as the reflection coefficient) is usually set to 1.0 for {\it
radiative} stars, and 0.5 for {\it convective} stars which transport some
of the incident energy to other regions of the star before re-radiating it
(Rucinski\markcite{29} 1969). The dependence of the bolometric flux on
local effective gravity is described by the gravity darkening exponent; a
value of 1.0 corresponds to a direct proportionality between flux and
gravity, and is appropriate for {\it radiative} stars (von~Zeipel
\markcite{30}1924).  The flux at any point on the surface of a {\it
convective} star is less dependent on the local gravity, and the exponent
is thought to be 0.32 (Lucy\markcite{31} 1967). Since the temperature of
BH~Cas is well within the convective regime, the gravity darkening
exponents and reflection coefficients should be near the theoretical
values for convective stars. These are the values that I assumed for all
of the modeling.

\subsection{GAWD Code}

In the summer of 1997, I began to develop a simple GA-based optimization
routine for the 1993 version of the \hbox{W-D} code ({\tt GAWD}). Inspired 
by a plot in a paper by Stagg \& Milone\markcite{32} (1993) which showed a
slice of ill-behaved parameter space in the $\Omega$-$q$ plane, I decided
to try using a GA to optimize in this two-dimensional space first. The
results of this initial test were encouraging. I used the \hbox{W-D} code to
generate a synthetic V--band light curve, and then I let {\tt GAWD} try to
find the original set of parameters from a uniform random initial sampling
of 1000 points in the surrounding region of $\Omega$-$q$ space. After 10
generations, nearly 99\% of the trial parameter-sets were statistically
indistinguishable from the original set of parameters.

The {\tt GAWD} code quite naturally divided into two basic functions:
using the \hbox{W-D} code to calculate light curves, and manipulating the 
results from each generation of trial parameter-sets. The majority of the
computing time is spent calculating the geometry of the binary system for
each set of model parameters. The GA is concerned only with collecting and
organizing the results of {\it many} of these models, so I incorporated
the message passing routines of the public-domain PVM software (Geist {\it
et al.}\markcite{33} 1994) to allow the execution of the code in parallel
on a network of 25 workstations.

With this pool of computational resources, the {\tt GAWD} code evolved to
do more than I originally thought would be feasible. Adding the capability
to fit for more than two parameters was simply a matter of extending the
GA to deal with longer strings of numbers. The light curve models take the
same amount of time to run regardless of the number of parameters which
are specified by the GA, so in some sense the extra parameters were added
for free. This is not strictly true since increasing the dimensionality of
the parameter space slows the convergence of the GA. After experimenting
in the simple $\Omega$-$q$ space for awhile, I added the capability to fit
for the inclination $i$, the temperature ratio of the two stars $T_1/T_2$,
and finally the temperature of the primary star $T_1$. Since the absolute
temperature of a star is unconstrained by observations in only one
bandpass, I altered {\tt GAWD} to fit light curves in the UBV bandpasses
simultaneously.

\subsection{Global Search\label{pssect}}

The first thing I had to do before starting {\tt GAWD} was specify the
ranges of the 5 parameters: (1) The measurement of the mass ratio
($q\equiv\frac{m_2}{m_1}$) from the radial velocity data placed strong
constraints on the allowed range.  Initially I fit a spectroscopic
orbit to the radial velocity data allowing the orbital period to be a free
parameter, and with the eccentricity fixed at 0.0. The period of the orbit
was consistent with the photometric period, so I fixed it for the final
fit. I allowed {\tt GAWD} to fit for a mass ratio between $\pm3\sigma$ of
the final spectroscopic value. (2) The shape of the observed light curves
indicates that BH~Cas is overcontact, so I constrained the equipotential
parameter $\Omega$ to be somewhere between the inner and outer critical
equipotentials for a given mass ratio. (3) The depth of the eclipses
implies that the inclination is fairly high, but I allowed all values
above $i=50\arcdeg$. (4) The temperature ratio is strongly constrained by
the relative depths of the two minima, so I included all values between
0.93 and 0.97. (5) The photometric colors indicate that the components of
BH~Cas are fairly cool, so I allowed temperatures for the primary star
between $T_1=4200$ and 5000~K.

After turning on the 25-host metacomputer with PVM, I started the GA
master program on one of the faster computers. The process begins by
reading the observational data into memory and assigning each light curve
equal weight. A randomly distributed initial set of 1000 trial
parameter-sets is generated, and each set of parameters is sent out to a
slave host. Each slave calculates theoretical UBV light curves for the
given set of parameters and returns the result to the master. Upon
receiving a set of light curves, the master process sends a new job to the
responding slave and computes the variance of the three calculated light
curves compared to the observed data. The fitness of the trial
parameter-set is determined from a simple average using the three
individual variances.

When the results of all 1000 trial parameter-sets are in hand, the master
process normalizes the fitnesses to the maximum in that generation. One
copy of the fittest trial parameter-set is passed to the next generation
unaltered, and the rest of the new population is drawn from the old one at
random, weighted by the fitness. Each trial is encoded into a long string
of numbers, which are then paired up for manipulation.  The single point
crossover operator is applied to 65\% of the pairs, and 0.3\% of the
encoded numbers are altered by the mutation operator.  The shuffled trial
parameter-sets are decoded, and those which are still within the allowed
ranges of parameters replace their predecessors. Computations of the new
set of models are distributed among the slave hosts, and the entire
process is repeated until the fractional difference between the average
fitness and the best fitness in the population is smaller than 1\%. As the
evolution progresses, correlations between parameters are revealed in the
spatial distribution of trial parameter-sets after the worst have been
eliminated, but before the final solution has converged. The parameters of
the best solution in the final generation provide the initial guess where
the traditional approach begins.

At this point, it is instructive to ask: has the algorithm found the
actual region of the global minimum? This is really an epistemological
question.  First, the algorithm performs a global random sampling of
parameter space. After the initial convergence, the crossover and mutation
operations continue to explore new regions of the parameter space in an
attempt to find better solutions. Repeating this procedure many times with
different random number seeds helps to ensure that the minimum found is
truly global; but no solution can, with absolute certainty, be proven to
be the global minimum unless every point in the parameter space is
explicitly evaluated.  At best, an algorithm can only take steps to sample
parameter space in a sufficiently comprehensive way so that the
probability of converging to the global minimum is very high.

The fitting routine supplied with the \hbox{W-D} code is a differential
corrections procedure. This program calculates light and radial velocity
curves based on the user-supplied first guesses for the model parameters,
and then recommends small changes to each parameter based on the local
shape of the parameter space. After the corrections have been applied, new
curves are calculated and the shape of the local parameter space is again
determined. New corrections are suggested, and the process continues until
the suggested corrections for all parameters are smaller than the
uncertainties.

\section{Results}

I evolved the `first guess' set of parameters for BH~Cas using the {\tt
GAWD} code. The GA randomly populated the parameter space defined in \S
\ref{pssect} and allowed the set of trial parameter-sets to evolve. After
100 generations, the difference between the average set of parameters and
the best set of parameters was insignificant. The parameter values in the
final generation of 1000 trial parameter-sets averaged to (mean $\pm\ 1
\sigma$):
\begin{eqnarray*}
q^{(1)}         &=&    0.474 \pm 0.002        \\
\Omega^{(1)}    &=&    2.798 \pm 0.015        \\
i^{(1)}         &=& 69\fdg52 \pm 1\fdg42      \\
\left[T_1/T_2\right]^{(1)} &=&    0.953 \pm 0.005        \\
T_1^{(1)}       &=&     4788 \pm 106\ {\rm K}
\end{eqnarray*}
The {\tt GAWD} code utilized only the light curve data to arrive at this
result.  All data points were included, and each light curve was assigned
equal weight in the assessment of fitness. The radial velocity data were
used only to restrict the range of possible mass ratios (see \S
\ref{pssect}).

For the final solution, I used the Wilson-Devinney differential
corrections code with all of the radial velocity data, and a subset of
$\sim$100 data points from each light curve. Starting with the best
solution from the {\tt GAWD} code, I allowed $\Omega$, $i$, $T_2$, and
$L_1$ to be free parameters. I iteratively applied any significant
corrections returned by the code to the parameters until all of the
corrections were small relative to their uncertainties.  Finally, I fixed
$T_2$ and $\Omega$ and allowed $T_1$ and $q$ to be free parameters
instead. No significant corrections were returned by the code, and the
final orbital solution for BH~Cas yielded:
\begin{eqnarray*}
q       &=&   0.474 \pm 0.002      \\
\Omega  &=&   2.801 \pm 0.003      \\
i       &=& 70\fdg1 \pm 0\fdg2     \\
T_1     &=&    4790 \pm 10~{\rm K} \\
T_2     &=&    4980 \pm 10~{\rm K}
\end{eqnarray*}
The uncertainties given here are probable errors from the DC output.
In Figure 2, the best-fit model from the \hbox{W-D} code is shown with 
the data for comparison. The deviations of the fit from the data may be 
the result ~of the blackbody ~assumption ~for ~the radiation ~law, or ~due ~to 
\newline\newline


\epsfxsize 3.5in
\epsffile{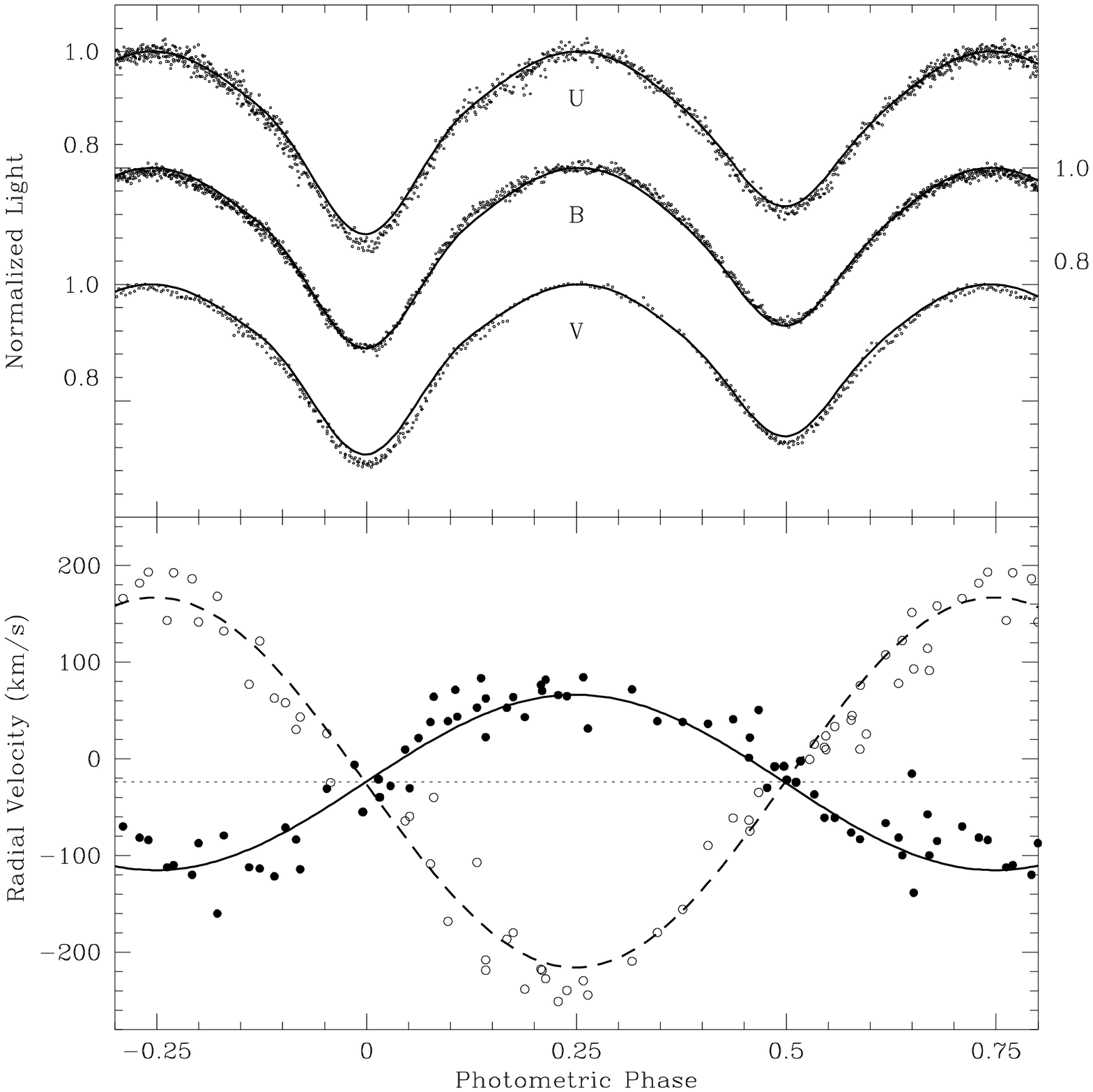}
\noindent{\sc Fig.~2.---} The best-fit model from the \hbox{W-D} code plotted
with the observational data that were used for the fit. In the top panel,
the observations are: U--band (upper), B--band (middle), and V--band
(lower). In the bottom panel, the observations are:  primary (solid
points) and secondary (open points) components of BH~Cas.



\begin{table*}
\tablenum{2}\label{tab2}
\centerline{TABLE 2}
\centerline{\sc Properties of the Eclipsing Binary BH Cassiopeiae}
\begin{center}
\vskip -5pt
\begin{tabular}{ll}\hline\hline
Binary Type \dotfill                          & W~UMa (W-type)            \\
Position $\alpha,\delta$ (J2000.0) \dotfill   & 00{\hbox{$^{\rm h}$}}21{\hbox{$^{\rm m}$}}21\fs41, +59\arcdeg09\arcmin05\farcs2\\
Proper motion $\mu_{\alpha},\mu_{\delta}$ (arcsec century$^{-1}$) \ldots & $-0.8\pm1.0, -2.2\pm1.0$  \\
Period (days) \dotfill                        & $0.405890(04\pm31)$       \\
Epoch of Primary Min. (JD$_{\sun}$) \dotfill  & $2449998.618(7\pm3)$      \\
Magnitudes at Max. $m_V,m_R,m_I$ \dotfill     & 12.58, 12.32, 11.67       \\
Spectral Type \dotfill                        & $\sim$K4$\pm$2            \\
Distance (pc) \dotfill                        & $\la 160$                 \\
Velocity amplitudes $K_1,K_2$ (km s$^{-1}$) \dotfill & $89.7\pm4.9,189.0\pm4.9$\\
Mass ratio $q$ \dotfill                       & $0.475\pm0.028$           \\
$\gamma$-velocity (km s$^{-1}$) \dotfill      & $-23.7\pm2.4$             \\
Space velocity (km s$^{-1}$) \dotfill         & $\la 29.6\pm4.9$          \\
Equipotentials $\Omega_1 = \Omega_2$ \dotfill & $2.801\pm0.015$           \\
Inclination $i$ \dotfill                      & $70\fdg1 \pm 1\fdg4$      \\
Temperatures $T_1, T_2$ (K) \dotfill          & $4790\pm100, 4980\pm100$  \\
Masses $m_1$,$m_2$ (M$_{\sun}$) \dotfill      & $0.74 \pm 0.06, 0.35 \pm 0.03$ \\
Pri. Radii $r_1({\rm pole},{\rm side},{\rm back})$ (R$_{\sun}$) \dotfill & $0.98\pm0.01, 1.05\pm0.01, 1.12\pm0.01$ \\
Sec. Radii $r_2({\rm pole},{\rm side},{\rm back})$ (R$_{\sun}$) \dotfill & $0.70\pm0.01, 0.73\pm0.01, 0.82\pm0.01$ \\
Pri. Fractional Lum. $L_1 (U,B,V)$ \dotfill   & $0.414\pm0.007, 0.399\pm0.005, 0.385\pm0.005$ \\
Sec. Fractional Lum. $L_2 (U,B,V)$ \dotfill   & $0.586\pm0.007, 0.601\pm0.006, 0.615\pm0.005$ \\
Limb darkening $x_1 = x_2$ $(U,B,V)$ \dotfill & 1.000, 0.976, 0.805       \\
Gravity darkening $g_1 = g_2$ \dotfill        & 0.32                      \\
Bolometric albedo $r_1 = r_2$ \dotfill        & 0.50                      \\
\hline
\end{tabular}
\end{center}
\end{table*}


\noindent the assumptions made during the reduction procedure, or both. 
Copies of the data shown in Figure 2 are available in the electronic edition 
of this paper.

I have listed the properties of BH~Cas (with $1\sigma$ errors) in Table
\ref{tab2}. The position and proper motion were obtained from a pair of
51-cm Astrograph plates on the measuring machine at Lick Observatory
(A.~Klemola, personal communication). The results were noted to be
somewhat worse than is normally expected from Astrograph plates due to the
weakness of the exposures and the location of the image near the plate
edge.

From 12 times of minimum light spanning nearly 2000 cycles (Metcalfe
\markcite{34}1997), I derived the following ephemeris for BH~Cas:
\begin{eqnarray*}
{\rm Min~I}&=&{\rm JD}_{\odot}~2449998.618(7 \pm 3) + \\
           & & +\ 0\fd405890(04 \pm 13)\times{\rm E} 
\end{eqnarray*}
The residuals of this fit exhibit no systematic trend, and the single
largest departure is $\sim1\times10^{-3}$ days.

I derived the magnitudes of BH~Cas at maximum light in the V--, R--, and
I--bands from CCD observations using the Selected Areas for zero-point
calibration (Landolt\markcite{35} 1992). The colors of W~UMa stars change
only slightly with orbital phase because the common envelope forces the
secondary star to be hotter than it would be if isolated. Observationally,
this appears to be true regardless of the mass ratio of the system (Shu \&
Lubow\markcite{36} 1981). The $V-I$ and $R-I$ colors imply an effective
temperature for BH~Cas near $T_{\rm eff} = 4600 \pm 400\ {\rm K}$,
corresponding to a Main Sequence spectral type of roughly K4$\pm$2
(Bessell\markcite{37} 1979).

An isolated Main Sequence star with this same effective temperature would
have an absolute visual magnitude $M_V = +7.0 \pm 0.7$ (Allen\markcite{d}
1973). When this is combined with the observed magnitude it provides an
upper limit on the distance. The primary (more massive, cooler) component
of BH~Cas has a structure which is approximately the same as an isolated
Main Sequence star of comparable mass. The same cannot be said of the
secondary (less massive, hotter) component. As a consequence, it is best
to use the observed magnitude at the orbital phase when essentially all of
the light is coming from the primary component. For BH~Cas, this occurs
during the minimum of the deeper eclipse when $m_V=13.1$. The resulting
upper limit on the distance is $d \la 160\ {\rm pc}$.

The mass ratio and center of mass radial velocity of the binary system
listed in Table \ref{tab2} comes directly from the spectroscopic orbital
solution. I obtained identical results whether or not the few observations
during eclipses were included in the fit. The transverse component of the
space velocity is determined by combining the distance with the measured
proper motion. The total proper motion is $\mu_{\rm tot} = (\mu_{\alpha}^2
+ \mu_{\delta}^2)^{1/2} = 2.34\pm1.0$ arcsec century$^{-1}$. At a distance
of 160 pc, this motion corresponds to a transverse velocity of $v_T =
17.7\pm7.6$ km s$^{-1}$. Combining this with the measured radial velocity
of the system, the total space velocity is $v_S \la 29.6\pm4.9$ km
s$^{-1}$.  Thus, BH~Cas appears to be a member of the disk population.

In addition to the mass ratio, the spectroscopic orbit also yields the
total mass of the binary system in terms of the semi-major axis, the
orbital period, and some fundamental constants. If the semi-major axis is
expressed in terms of the mass ratio, K amplitudes, and inclination, then
the individual masses can be calculated. I list the masses obtained by
using the best-fit value of the inclination.

Once the mass ratio is established and the scale of the system is
determined from the K amplitudes, the best-fit value for the equipotential
parameter leads directly to the absolute radii of the two stars. Since
BH~Cas is an overcontact system, the stars are tidally distorted, and no
single number can adequately define the radii; I list the radii at the
pole, side, and back.

\section{Discussion}

The orbital period of BH~Cas places it comfortably within the range of
periods typical of many W-type W~UMa systems. The mass ratio is near the
average for this class, but the individual masses are near the low end of
typical values. The mass of the secondary star seems particularly low, but
this is not uncommon among W~UMa systems. The secondary mass in V677~Cen,
for example, is 0.15 M$_{\sun}$ (Barone, Di Fiore \& Russo\markcite{38}
1993).  In a recent paper containing a large self-consistent sample of
absolute elements for W~UMa systems (Maceroni \& van't Veer\markcite{39}
1996), fully 12\% of the W-type, and 18\% of all W~UMa secondary stars had
masses less than or equal to the secondary mass derived for BH~Cas.  The
total system mass, however, is smaller than that of any W~UMa star in this
same sample, the smallest of which is CC~Com (1.20 M$_{\sun}$ total).

The absolute radii of the components of BH~Cas are larger than expected
for stars with such low masses. The average radii are similar to those of
TY~Boo, another W-type system which has roughly the same mass ratio as
BH~Cas, but a systematically higher total mass (Milone {\it et al.}
\markcite{40}1991).  There are a number of possibilities which could have
led to biased estimates of the masses of BH~Cas. I derived the velocities
exclusively from the H$\alpha$ line since it was the only strong feature
available. I used an unbroadened radial velocity standard for the cross
correlation template.  Also, I assumed Gaussian profiles for the
correlation peaks.

The fit to the radial velocity data could also be a problem. There is
considerable scatter in the observations, perhaps enough to accommodate
the increased K amplitudes that would be needed to make BH~Cas as massive
as TY~Boo. A set of high dispersion spectra of BH~Cas obtained at the
Multiple Mirror Telescope in 1994 has recently been re-analyzed using two
dimensional cross correlation techniques with synthetic templates
(G.~Torres, personal communication).  The re-analysis yielded a mass ratio
consistent with the value derived from the echelle spectra, but the
best-fit K amplitudes were considerably larger. The correlations which led
to these results, however, were very weak.

A significant x-ray signal at the position of BH~Cas was serendipitously
discovered by Brandt {\it et al.}\markcite{41} (1997). They found an x-ray
flux of $4.2\times10^{-14}$ erg cm$^{-2}$ s$^{-1}$. The implied upper
limit on the x-ray luminosity from the distance given above is $L_x \la
1.0\times10^{28}\ {\rm erg\, cm^{-2}\, s^{-1}}$. The detection of x-rays
from BH~Cas is not too surprising considering that most W~UMa systems show
clear signs of magnetic activity (Maceroni \& van't Veer\markcite{39}
1996). The mechanism for x-ray production is generally thought to be
related to magnetic reconnections in the corona (Priest, Parnell \& Martin
\markcite{42}1994). The value of the x-ray luminosity of BH~Cas is lower
by 1--2 orders of magnitude compared to other nearby W~UMa stars (McGale,
Pye \& Hodgkin\markcite{43} 1996).

\acknowledgements
I would like to thank Robert Jedicke and R.E.~White for their guidance
during the early stages of this project, Staszek Zola for sharing his
expertise in fitting light curves with the \hbox{W-D} code, and Ed Nather for
many helpful discussions. Finally, I wish to thank the anonymous referee
for a very helpful and thorough review which significantly improved the
manuscript. This work was supported in part by a fellowship from the Barry
M.~Goldwater Scholarship and Excellence in Education Foundation, an
undergraduate research grant from the Honors Center at the University of
Arizona, and grant AST-9315461 from the National Science Foundation.

\end{document}